\documentclass[aps,pre,twocolumn,showpacs,showkeys,floatfix]{revtex4}

\usepackage{graphicx}
\usepackage{dcolumn}
\usepackage{bm}

\begin{document}

\title{Computationally efficient optimization of radiation drives}

\date{September 12, 2015, revisions to March 11, 2019
   -- LLRL-JRNL-769474}

\author{Damian C. Swift}
\affiliation{%
   Lawrence Livermore National Laboratory,
   7000 East Avenue, Livermore, California 94551, USA
}
\author{George B. Zimmerman}
\affiliation{%
   Lawrence Livermore National Laboratory,
   7000 East Avenue, Livermore, California 94551, USA
}

\begin{abstract}
For many applications of pulsed radiation, the time-history
of the radiation intensity must be optimized to induce a desired
time-history of conditions.
This optimization is normally performed using multi-physics simulations 
of the system. The pulse shape is parametrized, and multiple
simulations are performed in which the parameters are adjusted until the
desired response is induced.
These simulations are often computationally intensive, and the optimization
by iteration of parameters in forward simulations is then expensive and slow.
In many cases, the desired response can be expressed such that an instantaneous
difference between the actual and desired response can be calculated.
In principle, a computer program used to perform the forward simulation
could be modified to adjust the instantaneous radiation drive automaticaly
until the desired instantaneous response is achieved.
Unfortunately, such modifications may be impracticable in a complicated
multi-physics program.
However, the computational time increment in such simulations is generally
much shorter than the time scale of changes in the desired response.
It is much more practicable to adjust the radiation source so that the response
tends toward the desired value at later times.
This relaxed in-situ optimization method can give an adequate design 
for a pulse shape in a single forward simulation,
giving a typical gain in computational efficiency of tens to thousands.
This approach was demonstrated for the design of laser pulse shapes
to induce ramp loading to high pressure in target assemblies
incorporating ablators of significantly different mechanical impedance 
than the sample, requiring complicated pulse shaping.
\end{abstract}


\maketitle

\section{Introduction}
For many applications of pulsed radiation, the time-history
of the radiation intensity must be optimized to induce a desired
time-history of conditions.
For example, temporal shaping of a laser pulse has been used to induce ablation
of a solid target, producing
a variety of mechanical loading histories in a solid sample,
in particular shock \cite{Swift_elements_2004}
and ramp compression \cite{Swift_lice_2005}.
The optimization of the radiation intensity to produce the desired effect 
is normally performed using multi-physics simulations of the system.
Each simulation solves an initial-value problem: the evolution equations
are integrated forward in time, using an estimate of the intensity
history of the radiation applied to the system \cite{radhydrosimex}.
Repeated simulations are performed, adjusting the pulse shape to be used in each
until the desired response is induced.
These simulations are often computationally intensive, and the optimization
by iteration of parameters in forward simulations is then expensive and slow.
The workload to a human designer may be reduced, and the accuracy of the design
improved, by wrapping the forward simulation in an automated optimizer
employing a numerical search strategy \cite{iterativefwd},
but the computational effort is often greater.

In this paper, we discuss an alternative approach that can be applied
when the state at some location within the experimental assembly
responds rapidly to changes in the radiation intensity,
by modifying the intensity automatically during the simulation.

\section{Modifications to radiation hydrocode for pulse shape design}
In a range of useful applications, including the design of radiation
intensities for the generation of mechanical or thermal loading,
the state in the sample responds promptly to changes in the intensity.
This is the case for the pressure at the critical surface in laser ablation.
For a close coupling between applied intensity and resulting response,
the multi-physics forward simulation could in principle be modified to
adjust the radiation intensity at each instant of time until the desired
response is produced.
This modification is relatively invasive, requiring significant changes
to the structure of the computer program used to perform the simulation.
In contrast, an iterative wrapper can be constructed 
for a forward simulation without any modifications to the simulation program.
However, there is an enormous potential advantage in that the intensity
history of the radiation pulse could be calculated in a single simulation.

In other applications, the state in the sample does not respond so promptly
to changes in the radiation intensity, but it may be related more simply
to states induced in a region which does respond promptly.
For instance, if an ablator \cite{ablator,ablator2}
is used to apply a load to a sample in contact with the ablator,
the pressure history in the sample can often be related to the pressure history
at the critical surface of the ablator using continuum dynamics without
radiation transport \cite{Swift_alloys_2004}.
In the absence of radiation transport, hydrocode simulations require much
less computational effort, and may be accomplished by different means such as 
the backward propagation of characteristics \cite{backwards}, 
allowing a desired loading history in the sample
to be transformed to a desired pressure history at the ablation surface.
Our procedure can then be used to determine the radiation intensity
from this modified state history.

Ideally, a single time increment of the radiation hydrocode would be
repeated, adjusting the intensity of the radiation source until the desired
response is achieved.
In practice, this is a relatively complicated modification to a
general-purpose, multi-physics computer program.
A simpler approach is to perform a sequence of time increments as usual,
adjusting the radiation source such that the response becomes closer to
the desired response.
Usually, the time increments in the simulation are much shorter than 
the time scale of the
desired response, so some degree of lag in the induced response is
acceptable.
It is difficult to devise a general algorithm to adjust the source
intensity so that the response approaches the instantaneous desired value
rapidly and monotonically.
In practice, reasonable results can be obtained with forward-time
schemes which are not completely stable -- the induced response, 
and thus the radiation intensity, may oscillate about the desired history.
Conversely, the radiation intensity may respond too slowly, so the approach to
the desired response is over-damped.
However, a reasonable radiation intensity history may be found
by inspection of the calculated intensity history and response history.
Improved results can then be obtained by repeating the simulation
a small number of times, using the calculated radiation intensity history
from the previous simulation, or a smoothed version.

\section{Examples for loading histories induced by laser ablation}
The main application motivating this study was to make the design
of laser pulse shapes for dynamic loading experiments more efficient.
Laser systems are notable in that the power history can be controlled
relatively easily
to produce highly-structured shapes, and the pressure response
of the ablation plasma is relatively rapid.
In contrast, the temperature history of radiation within a hohlraum
and the current history in an electrical pulsed power discharge can typically
not be controlled with the same degree of flexibility.

Radiation hydrodynamics simulations were performed using the 
{\sc Lasnex}\cite{lasnex}
and {\sc Hydra} \cite{hydra} computer programs.
The current incarnation of {\sc Lasnex} was written using the interpreted
language {\sc Basis} \cite{basis} in the user interface. Pulse-shaping
algorithms were written through the {\sc Basis} input without requiring any
modification to the {\sc Lasnex} source code.
Similarly, {\sc Hydra} includes an interface to a Python interpreter,
which was used to implement the pulse-shaping algorithms through input files,
without changing the {\sc Hydra} source code.

The desired response to be controlled by the laser pulse shape was the
ablation pressure $p_a$, to reproduce some desired history $p_d(t)$.
As material is ablated, the location of the pressure-generating region
changes with respect to both the frame of reference of the undisturbed material
and also the original location of the material (i.e. in Eulerian and Lagrangian
frames).
An automated way is needed to locate the ablation region at any instant of time.
There are many possible ways to do this, with applicability to different
classes of problem.
Laser energy is deposited primarily around the critical surface.
We found reasonably general metrics to be the region over which the
mass density $\rho$ of ablated material drops by a few tens of percent
from its initial value.
For high-pressure loading, material about to be ablated may have been
compressed to significantly higher density, so this metric does not 
precisely identify the ablation front, but in many cases it locates the
critical surface to an adequate accuracy.
A higher reference density, representative of the desired loading pressure,
can give better results.
The instantaneous ablation pressure was estimated as the average pressure
over the ablation region identified in this way.

For applications where the desired pressure history increases monotonically
in time, an alternative metric is the maximum pressure in the problem.
If the simulation includes other materials, where the impedance mismatch
induces a higher pressure, this metric is not appropriate.
Also, if the power-adjusting algorithm induces a higher pressure than the
desired value, it may take significantly longer for the propagating
pressure wave to decay than for the ablation pressure to change,
so the power-adjusting algorithm may over-compensate for instantaneous 
discrepancies in ablation pressure.

Ideally, the power would be adjusted at each instant of time until the
desired ablation pressure is obtained.
This could involve repeatedly integrating the radiation hydrodynamics
equations over the same time increment, which would be complicated in
multi-physics codes such as {\sc Lasnex} and {\sc Hydra}.
The approach tried here was to adjust the laser power in the right direction to
correct $p_a$ toward $p_d$.
Adjustments were made with respect to a reference laser pulse shape $I_r(t)$,
using a scaling factor $\sigma(t)$.
An advantage of this approach is that multiple passes of the pulse-shaping
algorithm can be used, by taking the results of a previous simulation
as the reference pulse.
A wide range of strategies could be used to determine the instantaneous
value of $\sigma$ given $p_a$, $p_d$, and $I_r$.
The loading induced in matter by ablation is often approximated roughly by an
irradiance-pressure relation such as
\begin{equation}
p_a = \alpha I^\beta,
\end{equation}
where $\beta\sim$0.6-0.8 \cite{Swift_elements_2004}.
Thus, given the current irradiance $I(t)=\sigma(t)I_r(t)$,
a modified irradiance is predicted 
\begin{equation}
\tilde I = I\left(\frac{p_d}{p_a}\right)^{1/\beta},
\end{equation}
allowing a modified scaling $\tilde\sigma$ to be calculated.
In typical simulations, an unstable oscillatory irradiance was produced
if $\tilde\sigma$ was used directly for the next time step.
Instead, under-relaxation was used to vary the irradiance over time
scales over which laser pulse shapes can typically be controlled,
\begin{equation}
\sigma(t+\delta t)=\sigma(t)+\gamma\left(\tilde\sigma(t)-\sigma(t)\right)
\end{equation}
where $\gamma$ is a relaxation parameter $\sim 10^{-3}$.

Variants of this scheme were tried, such as limiting the absolute or fractional
rate of change of intensity.
This mimics the constraints of real lasers (and other sources of pulsed
energy), which typically have a finite response time with which pulse shapes
can be defined, limited usually by the bandwidth of the electronic
and optical modulator components used to control the gain of the amplifiers.
In many cases of interest, including pulse-shaping to produce a constant
shock pressure, an increasing ramp,
or a shock followed by a second shock or a ramp,
experiments and simulations show that the irradiance increases monotonically.
It was found that oscillatory behavior was reduced by constraining the
irradiance to be monotonic, at the expense of tending to exceed the target
pressure slightly.
The pulse shape could then be improved by performing a further iteration of the
algorithm.

Most of the simulations below were performed in one spatial dimension,
appropriate when the diameter of the laser spot is large enough that
lateral release in the ablation plume does not affect the center of the 
spot.
This constraint is typically more severe than for loading in the condensed
target to remain one dimensional, because the speed of sound in the ablaton
plasma is typically higher than in the condensed target.
However, the approach described can be applied equally well to
simulations in more than one space dimension,
as long as the critical surface can be identified automatically.
The advantage in efficiency is then correspondingly greater, although the
slower rate at which time increments occur mean that it takes longer to
ensure that a given numerical optimization strategy is behaving acceptably.

The optimized pulse shape did not depend on the initial guess chosen,
indicating that the solution was unique.
The choice of initial guess did affect the quality of the pulse shape
resulting from the first iteration, and the number of iterations needed to
give a pulse shape of given maximum deviation from the desired
pressure history.

\subsection{Constant ablation pressure}
A very common requirement in dynamic loading experiments is to induce
a constant ablation pressure, with as fast a rise as possible.
Usually, a constant ablation pressure requires a rising irradiance,
as the critical surface accelerates away from the surface and the mass
density there decreases.
If the duration is long enough with respect to the diameter of the drive spot,
lateral flow in the ablation plume must also be compensated by increasing
the irradiance.
Here we consider the case of a large drive spot, so 1D simulations are
adequate.
The irradiance history was calculated to induce 100\,GPa in Al for 10\,ns,
following a rise from zero over 200\,ps, as is common for high energy
laser systems.
The Al was 500\,$\mu$m thick, so release from the rear surface did not
affect the ablation region, which would complicate the pulse-shaping.

The simulations were performed with the {\sc Lasnex} program, version 1604081910.
Al was modeled using LEOS 130 \cite{LEOS} and OPAL opacities \cite{OPAL}.
Thomas-Fermi ionization was used.
The electron flux was limited to 0.05 of the free stream value,
a common choice for such simulations \cite{Dendy1993}.
As is necessary for simulations of laser ablation, the spatial zoning
varied exponentially from the free surface of the ablator, so that the
optical skin depth was properly resolved \cite{Swift_elements_2004}.

The optimization parameters chosen were $\beta=0.8$, $\gamma=2.5\times 10^{-3}$.
Optimization was made less sensitive while the laser pulse was coming up
to power, to allow the ablation plume to establish itself without trying to
follow any detailed pressure history.
The ablation region was identified as that where $0.5 < \rho/\rho_0 < 0.99$,
with $\rho_0=2.7$\,g/cm$^3$ for Al.
The initial guess for the irradiance was a constant 1\,TW/cm$^2$.

\begin{figure}
\begin{center}\includegraphics[scale=0.18]{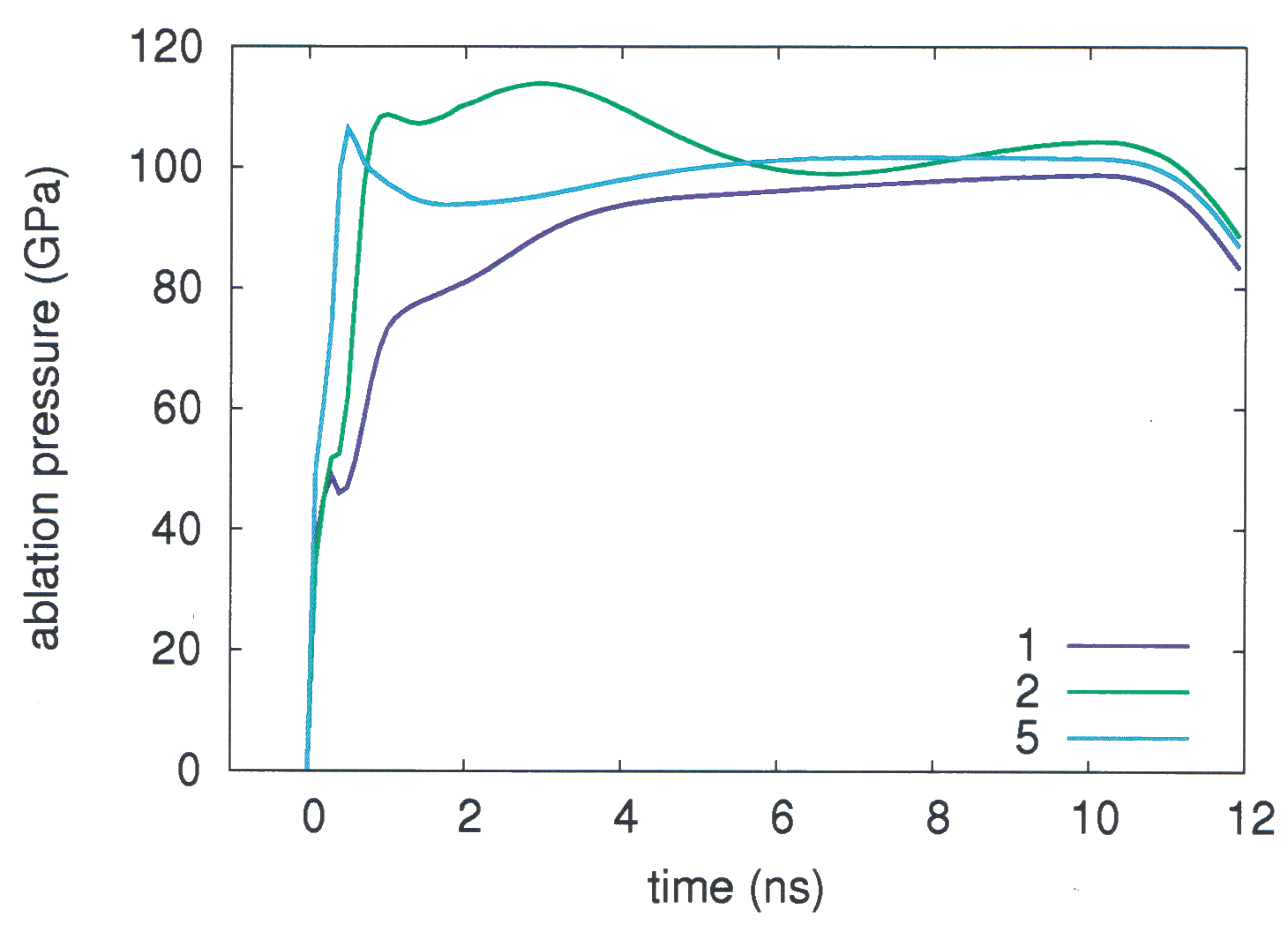}\end{center}
\caption{Pressure histories for design of constant-pressure shock in Al.}
\label{fig:shockp}
\end{figure}

\begin{figure}
\begin{center}\includegraphics[scale=0.18]{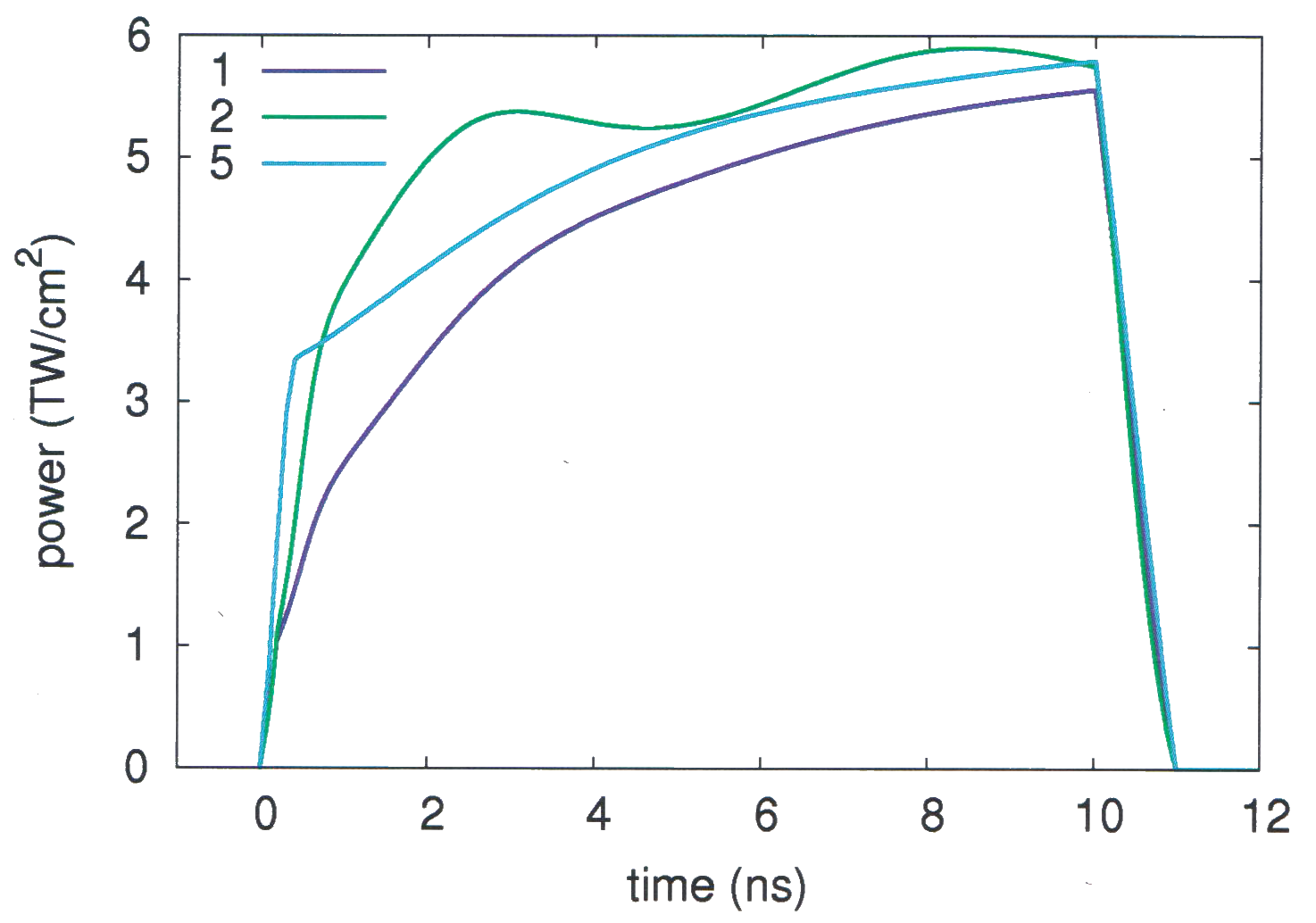}\end{center}
\caption{Irradiance histories for design of constant-pressure shock in Al.}
\label{fig:shocki}
\end{figure}

With the parameters chosen, a singe pass of
the pulse-shaping algorithm gave a pressure history that reproduced the
desired history to around 10\%, and to better than 10\%\ toward the
end of the pulse.
By performing a small number of iterations, the pressure could be
brought to within a few percent of the desired history over almost all of the
pulse.
(Figs~\ref{fig:shockp} and \ref{fig:shocki}.)

\subsection{Ramp loading}
Another common requirement is to induce a monotonically-increasing pressure
such that a shock does not form as the ramp propagates through some finite
thickness of the target \cite{Swift_lice_2005,Swift_idealramp_2008}.
Approaching the limits of irradiance and pulse length for a given laser,
shocks can form for relatively modest imperfections in the pulse shape
\cite{Eggert2009}, so it may be crucial to follow the ideal pulse shape
\cite{Swift_idealramp_2008} as accurately as possible.

\begin{figure}
\begin{center}\includegraphics[scale=0.18]{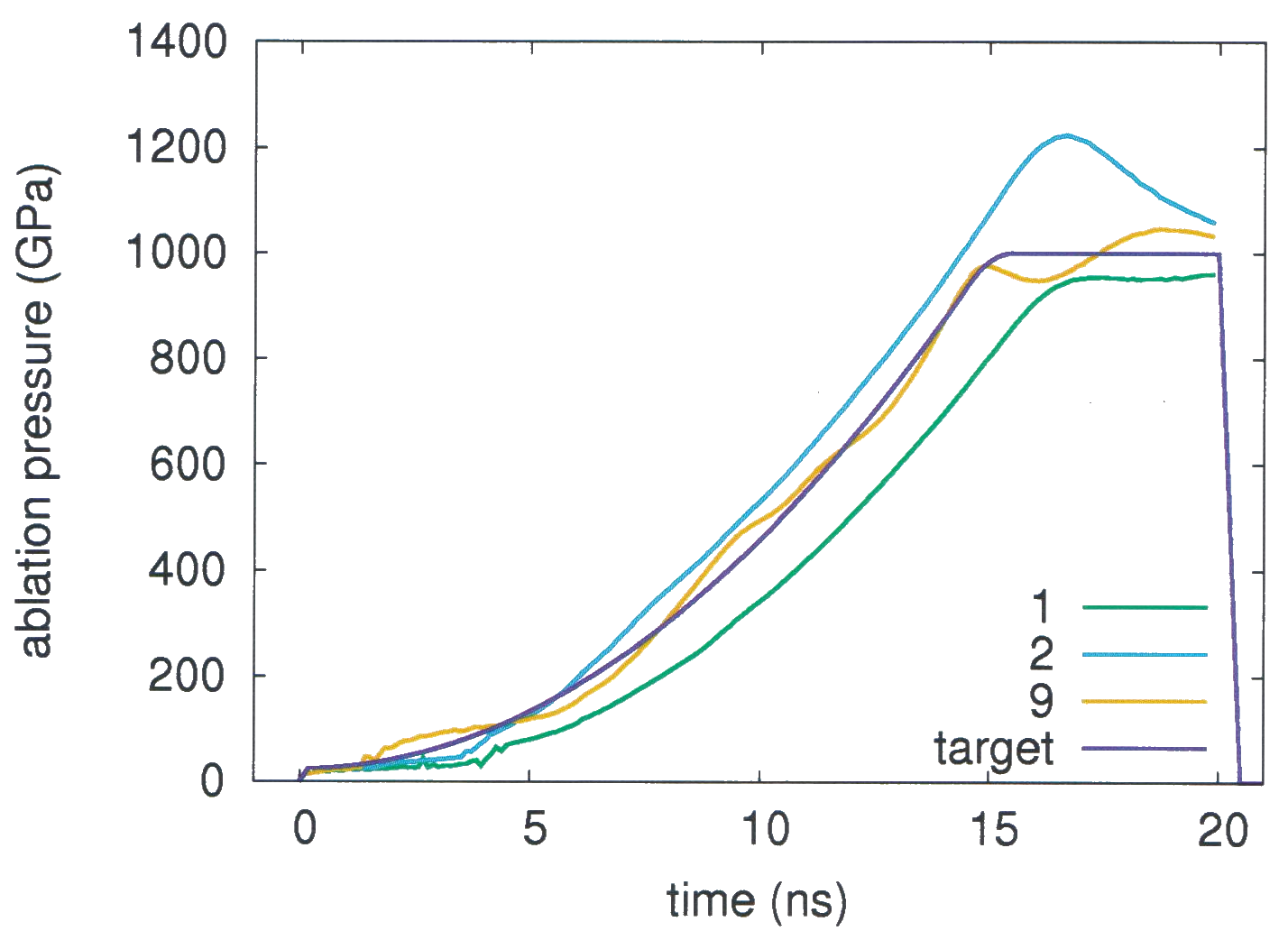}\end{center}
\caption{Pressure histories for design of ramp-hold in C.}
\label{fig:rampp}
\end{figure}

\begin{figure}
\begin{center}\includegraphics[scale=0.18]{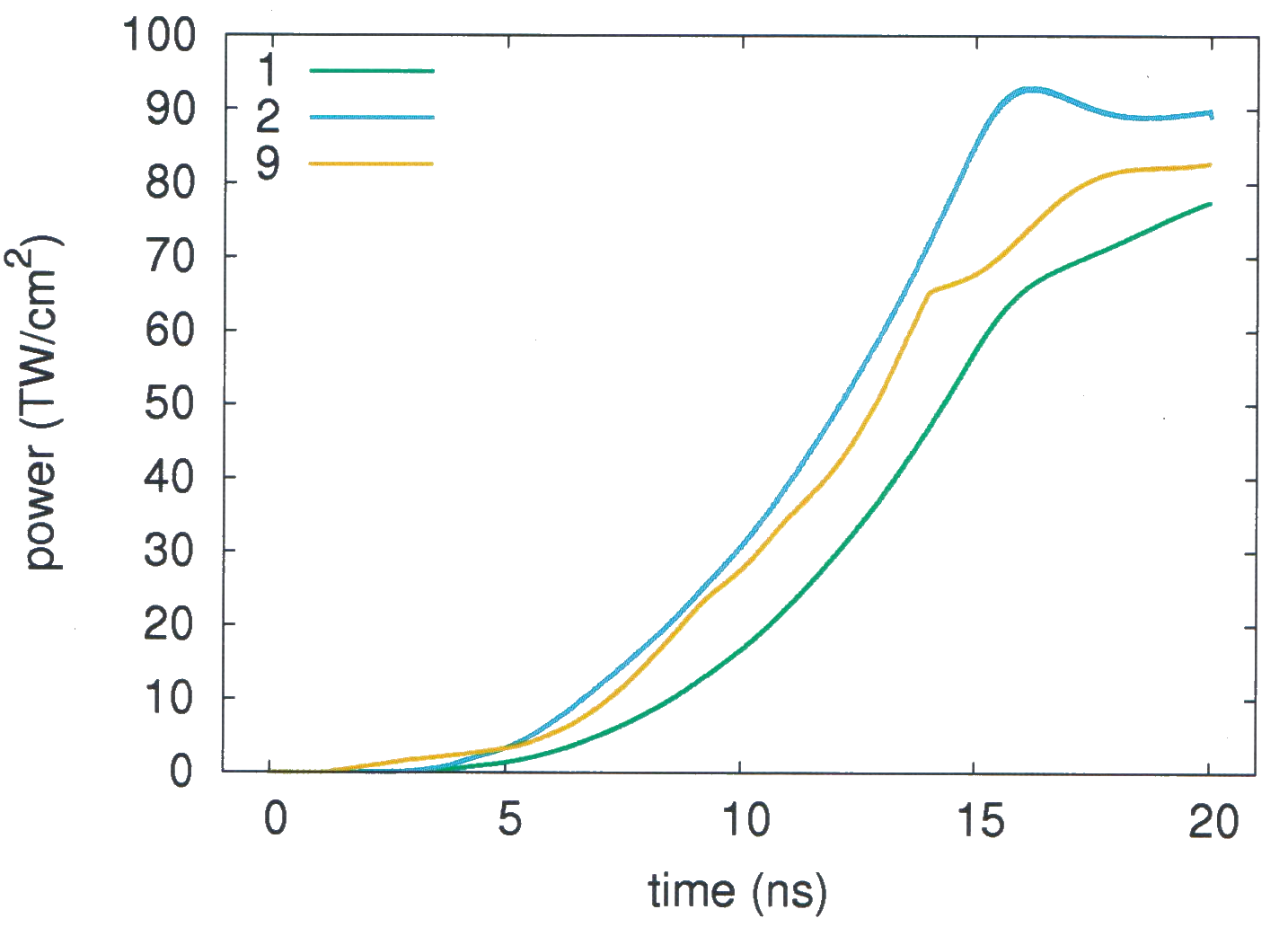}\end{center}
\caption{Irradiance histories for design of ramp-hold in C.}
\label{fig:rampi}
\end{figure}

Here we simulate the pulse shape needed to give the ideal ramp shape
to 1000\,GPa over 15\,ns in a diamond ablator,
followed by a constant ablation pressure for 5\,ns.
The diamond was modeled using an early version of a multiphase EOS
constructed using electronic structure calculations \cite{Correa2008},
an empirical strength model including electronic structure calculations
of shear modulus \cite{Orlikowski2007},
and OPAL opacities.
A single pass of the algorithm gave a pulse shape which reproduced the
desired pressure to 200\,GPa during the ramp, but induced oscillations
during the pressure hold.
Again, a small number of passes of the algorithm reproduced the desired
pressure history to arbitrary accuracy.
The algorithm was very stable during the ramp, but changes in the optimizer
parameters were needed for stability during the pressure hold.
(Figs~\ref{fig:rampp} and \ref{fig:rampi}.)

On further investigation, including the other pulse shapes described below,
oscillations tended to be triggered by the transition from one type of
loading to another, here from ramp to constant.
Different choices for the numerical optimization parameters were found to
make the irradiance and pressure more stable, but we did not find a universal
prescription.
At any instant of time,
the plasma plume reflects the loading history induced up to
that point. We hypothesize that the plume must in effect be reconfigured
for the change in loading history, and this cannot be achieved
instantaneously; attempting to do so induces oscillations.
Thus, in order to induce a ramp followed by a constant pressure, 
near the peak of the ramp, the pulse shape should anticipate the peak pressure
by reducing the loading rate gradually, rather than having it fall 
instantaneously to zero.
In practice, it was not necessary to guess a suitable target pressure history
in detail to achieve this effect.
Either the optimization parameters could be adjusted around the time of the
change to be less sensitive to pressure discrepancies and thus avoid
`hunting' for the correct solution, introducing oscillations which could grow,
or an improved pulse shape could be estimated by eye from the average baseline
around any oscillations, and used as a closer estimate for the next iteration.

\subsection{Shock followed by ramp}
Another useful loading path is to induce a steady shock, followed some time 
later by a ramp to a higher pressure, which is then sustained for some finite
time.
This loading history has been used to study solidification from the liquid,
where the shock is strong enough to melt the sample, and the ramp is strong
enough to pass back through the melt curve and into the solid
\cite{McNaney2009,Kraus2015}.
Shock-ramp loading can also be used when a laser system does not allow
long enough pulses to induce a ramp from zero pressure:
if weak enough, the entropy of the shock may be a small perturbation from
the isentrope in return for a significantly shorter overall pulse
\cite{Lazicki2015}.

\begin{figure}
\begin{center}\includegraphics[scale=0.18]{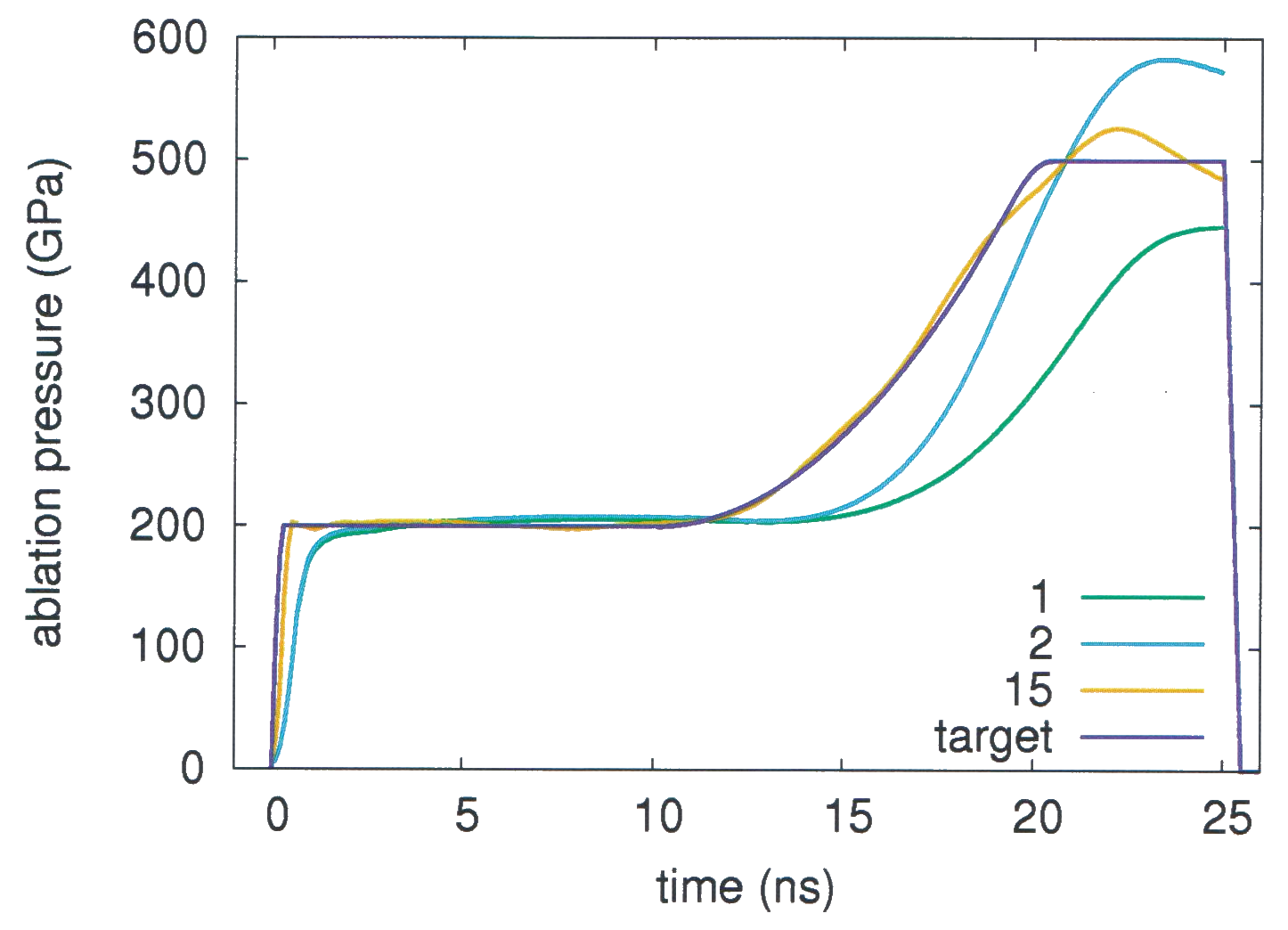}\end{center}
\caption{Pressure histories for design of shock-ramp-hold in kapton.}
\label{fig:shockrampp}
\end{figure}

\begin{figure}
\begin{center}\includegraphics[scale=0.33]{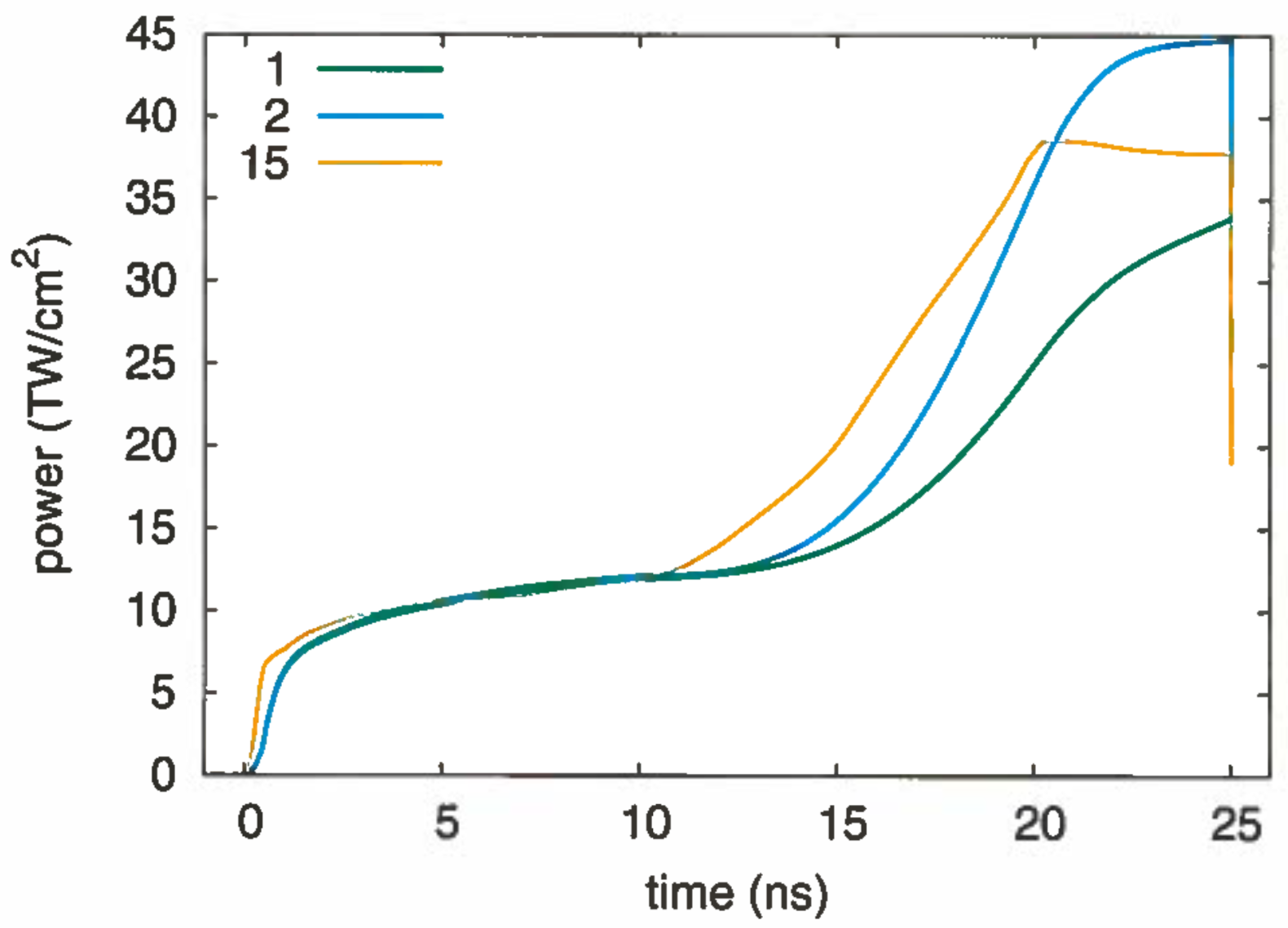}\end{center}
\caption{Irradiance histories for design of shock-ramp-hold in kapton.}
\label{fig:shockrampi}
\end{figure}

Here we simulate the pulse shape needed to give a shock to 200\,GPa, supported
for 10\,ns, followed by a ramp to 500\,GPa over 10\,ns, sustained for
5\,ns, in a kapton ablator.
The kapton was modeled using LEOS 5040 and OPAL opacities.
The first pass of the algorithm gave a pulse shape which reproduced the
desired pressure of the first shock, but induced strong oscillations during the
ramp drive.
Three further iterations with different optimization parameters were needed to
define the pulse shape for the ramp and sustained peak pressure.
A few additional calculations were performed to select the parameters.
(Figs~\ref{fig:shockrampp} and \ref{fig:shockrampi}.)

We observed that, if the ramp was sufficiently steep, the simulations
predicted that a second critical surface could form downstream of the
original one, and initially decoupled from the first.
The ramp in the sample was then delayed significantly until the
critical surfaces merged, significantly perturbing the desired pressure
history. The irradiance, adjusted automatically to try to induce the
desired pressure history, could then become incompatible with the
conditions in the plasma plume, inducing oscillations much as were observed
at the constant pressure peak following a ramp.
There seems therefore to be a maximum rate at which the nature of the pulse
shape should be changed, depending on the instantaneous plasma conditions:
a constraint to consider when designing pulse shapes.

\section{Conclusions}
An alternative approach to iterative forward optimization was investigated
for the design of radiation pulses to produce
a desired response in a target, by modifying the simulation program
to adjust the irradiance automatically during a simulation such that
the response approached the desired history.
This approach was investigated in the design of laser pulses to induce
a variety of loading histories for experiments on the response of condensed
matter to dynamic loading.
Existing multi-physics simulation programs were used, and the irradiance was
adjusted such that the load induced at the critical surface tended toward
the instantaneously-desired value, rather than inducing the desired value
at all times, which would have required significant modification to these
complicated programs.

Several optimization strategies were investigated, i.e. algorithms for
adjusting the irradiance given the instantaneous discrepancy between
calculated and desired ablation pressure.
In some cases, an acceptable pulse shape was found in a single
such simulation. More commonly, the induced pressure history exhibited
significant discrepancies and oscillations, and a small number of 
irradiance-adjusting simulations were needed, each using the output of the
previous simulation as an initial guess, with smoothing through oscillations
if of large amplitude.

It was found possible to induce oscillatory modes in the plasma that seemed
to be more than simply naive optimization strategies that overshot the
desired conditions, though the choice of optimization parameters could
certainly make the oscillations stronger and unstable.
We hypothesize an effective `memory' in the ablation plume, making it
necessary to change gradually between loading conditions of different
nature. In the case of ramp loading following a previously induced
constant (elevated) pressure, the initial ramp loading rate must be
slow enough to avoid the development of a second critical surface downstream
of the original one.

Overall, the use of this approach seemed to
significantly reduce the human and computational effort required to
design pulse shapes for laser loading experiments.

\section*{Acknowledgments}
David Munro provided advice on the {\sc Yorick} post-processing program
\cite{Munro1995}
for the {\sc Lasnex} radiation hydrocode.
Joe Koning provided advice on the interface to the
Python language within the {\sc Hydra} radiation hydrocode.
This work was performed under the auspices of
the U.S. Department of Energy under contract DE-AC52-07NA27344.

\end{document}